\documentstyle{article}
\def\1ad{\mbox{\normalsize $^1$}}
\def\2ad{\mbox{\normalsize $^2$}}
\def\3ad{\mbox{\normalsize $^3$}}
\def\4ad{\mbox{\normalsize $^4$}}
\def\5ad{\mbox{\normalsize $^5$}}
\def\6ad{\mbox{\normalsize $^6$}}
\def\7ad{\mbox{\normalsize $^7$}}
\def\8ad{\mbox{\normalsize $^8$}}
\def\makefront{\vspace*{1cm}\begin{center}
\def\newtitleline{\\ \vskip 5pt}
{\Large\bf\titleline}\\
\vskip 1truecm
{\large\bf\authors}\\
\vskip 5truemm
\addresses
\end{center}
\vskip 1truecm
{\bf Abstract:}
\abstracttext
\vskip 1truecm}
\newcommand{\pp}{{=\!\!\!|}}
\begin {document}
{\hfill{BRX-TH-423, VUB/TENA/98/1, hep-th/9801080}}
\def\titleline{
%%%%%%%%%%%%%%%%%%%%%%%%%%%%%%%%%%%%%%%%%%%%%%
%                                            %
% Insert now the text of your title.         %
% Make a linebreak in the title with         %
%                                            %
%            \newtitleline                   %
%                                            %
%%%%%%%%%%%%%%%%%%%%%%%%%%%%%%%%%%%%%%%%%%%%%%
Some aspects of $N=(2,2)$, $D=2$ supersymmetry\footnote{To appear 
in the proceedings of {\it Quantum aspects of 
gauge theories, supersymmetry and unification}, 
Neuchatel University, 18-23 September 1997.} 
%%%%%%%%%%%%%%%%%%%%%%%%%%%%%%%%%%%%%%%%%%%%%%
}
\def\authors{
%%%%%%%%%%%%%%%%%%%%%%%%%%%%%%%%%%%%%%%%%%%%%%
%                                            %
%  Insert now the name (names) of the author %
%  (authors).                                %
%  In the case of several authors with       %
%  different addresses use e.g.              %
%                                            %
%             \1ad  , \2ad  etc.             %
%                                            %
%  to indicate that a given author has the   %
%  address number 1 , 2 , etc.               %
%                                            %
%%%%%%%%%%%%%%%%%%%%%%%%%%%%%%%%%%%%%%%%%%%%%%
M.T. Grisaru \1ad , M. Massar \2ad, A. Sevrin \2ad, J. Troost \2ad \3ad
%%%%%%%%%%%%%%%%%%%%%%%%%%%%%%%%%%%%%%%%%%%%%
}
\def\addresses{
%%%%%%%%%%%%%%%%%%%%%%%%%%%%%%%%%%%%%%%%%%%%%%
%                                            %
% List now the address. In the case of       %
% several addresses list them by numbers     %
% using e.g.                                 %
%                                            %
%             \1ad , \2ad   etc.             %
%                                            %
% to numerate address 1 , 2 , etc.           %
%                                            %
%%%%%%%%%%%%%%%%%%%%%%%%%%%%%%%%%%%%%%%%%%%%%%
\1ad
Physics Department, Brandeis University, \\
Waltham, MA, USA\\
\2ad
Theoretische Natuurkunde, Vrije Universiteit Brussel, \\
Pleinlaan 2, B-1050 Brussel, Belgium\\
\3ad
Aspirant FWO
%%%%%%%%%%%%%%%%%%%%%%%%%%%%%%%%%%%%%%%%%%%%%%%
}
\def\abstracttext{
%%%%%%%%%%%%%%%%%%%%%%%%%%%%%%%%%%%%%%%%%%%%%%%
%                                             %
% Insert now the text of your abstract.       %
%                                             %
%%%%%%%%%%%%%%%%%%%%%%%%%%%%%%%%%%%%%%%%%%%%%%%
The off-shell description of $(2,2)$ supersymmetric non-linear
$\sigma$-models is reviewed. The conditions for ultra-violet finiteness
are derived and T-duality is discussed in detail.
%%%%%%%%%%%%%%%%%%%%%%%%%%%%%%%%%%%%%%%%%%%%%%%%%%%%%%%%%%%%%%%
}
\makefront
%%%%%%%%%%%%%%%%%%%%%%%%%%%%%%%%%%%%%%%%%%%%%%%%
%                                              %
%  Insert now the remaining parts of           %
%  your article.                               %
%                                              %
%%%%%%%%%%%%%%%%%%%%%%%%%%%%%%%%%%%%%%%%%%%%%%%%
\section{Introduction}
The close relation between supersymmetry and complex geometry becomes
particularly rich
in two dimensions. We focus on
non-linear $\sigma$-models with $N=(2,2)$ supersymmetry. These are not only
building blocks for type $II$ superstrings, they describe
the matter sector of $N=(2,2)$ string theories as well.

Any bosonic non-linear $\sigma$-model can be extended to an $N=(1,1)$
supersymmetric model. The model is completely
determined once a target manifold, ${\cal M}$, its metric, $g$,
and a closed 3-form, $T$, are specified \cite{GHR}.
Classically, $(2,2)$ supersymmetry requires the existence of two covariantly
constant complex structures, which are such that the metric is hermitean w.r.t.
both.
These models do have an off-shell
description which becomes manifest in (2,2) superspace. Such a formulation
yields the surprising result that the local geometry of all manifolds
which allow for (2,2) supersymmetry is encoded in a potential, the
Lagrange density. This generalizes the most important property of K\"ahler
manifolds to a large class of complex manifolds. The superspace
description also facilitates the study of the ultra-violet properties of
these models. Finally $T$-duality transformations can be studied while
keeping the $(2,2)$ supersymmetry manifest.
The present paper reviews the off-shell description of $N=(2,2)$ non-linear
$\sigma$-models. Subsequently, we study their ultra-violet properties and
finally the classical, abelian $T$-duality transformations are presented.
\section{$N=(2,2)$ non-linear $\sigma$-models}
As already mentioned, $N=(2,2)$ supersymmetry for a $\sigma$-model in $(1,1)$
superspace is equivalent to the
existence of two complex structures $J_+$ and $J_-$, which are
covariantly constant, $\nabla_c^+J_+^a{}_b=\nabla_c^- J_-^a{}_b=0$, where
$\nabla^\pm$ denotes covariant differentiation using the $ \Gamma^a_{\pm
bc}\equiv \left\{ {}^{\, a}_{bc} \right\} \pm T^a{}_{bc}$ connection, and which
are
such that the metric is hermitean w.r.t. both of them, $J_{\pm \,
ab}=-J_{\pm\, ba}$. On-shell, one gets the standard $N=(2,2)$
supersymmetry algebra, while off-shell closure is only achieved along
$\ker [J_+, J_-]=\ker(J_+-J_-)\oplus \ker(J_++J_-)$.

Building on the results of \cite{martinnew}, it was argued in
\cite{icproc} that, in order to achieve a manifest $(2,2)$ supersymmetric
description of these models, {\it i.e.} a formulation in $(2,2)$
superspace, chiral, twisted chiral \cite{GHR} and semi-chiral \cite{BLR}
fields are sufficient. More explicitly: the tangent space at any point of
the target manifold can be decomposed into three subspaces: $\ker (J_+ -
J_-)\oplus \ker (J_++J_-) \oplus (\ker[J_+, J_-])^\perp $. These
subspaces are conjectured to be integrable to chiral, twisted chiral and
semi-chiral coordinates respectively.

In order to obtain the off-shell formulation of these models, we describe it in
$N=(2,2)$
superspace. We call its coordinates $x^\pp$,
$x^=$, $\theta^+$, $\theta^{\dot +}$, $\theta^-$ and $\theta^{\dot
-}$. The fermionic derivatives satisfy $\{D_+,D_{\dot
+}\}=\partial_\pp$ and $\{D_-,D_{\dot
-}\}=\partial_=$, with all other (anti-)commutators vanishing. The $N=(1,1)$
fermionic
coordinates are the real part of the $N=(2,2)$ fermionic coordinates.
The non-chiral measure $d^2xd^4\theta$ is dimensionless, so
the Lagrange density is necessarily a function of scalar superfields. This
implies that all
dynamics arises from imposing constraints on general superfields. The
simplest constraints are linear in the derivatives, and they turn out to
be sufficient for our present purpose. It can be shown \cite{icproc}
that through an appropriate
coordinate transformation, this reduces to the three standard types of
superfields:

\vspace{.1cm}

\noindent{\it i.} \underline{Chiral superfields}: $z^a$, $\bar z^{\bar a}$;
$a,\,\bar a\in
\{1,\cdots d_c\}$.\\
$D_+ z^a= D_- z^a=D_{\dot +} \bar z^{\bar a} = D_{\dot {-}} \bar z^{\bar a} =
0$.\\
{\it ii.} \underline{Twisted chiral superfields}: $w^m$, $\bar w^{\bar m}$;
$m,\,\bar m\in
\{1,\cdots d_t\}$\\
$D_+ w^m=  D_{\dot -} w^m= D_{\dot +}\bar w^{\bar m}=D_- \bar w^{\bar m} =0$.\\
{\it iii.} \underline{Semi-chiral superfields}: $ r^\alpha $, $\bar
r^{\bar\alpha }$,
$ s^{\tilde\alpha} $, $\bar s^{\tilde{\bar\alpha} }$;
$\alpha ,\, \bar\alpha ,\, \tilde\alpha ,\, \tilde{\bar\alpha}\in\{
1,\cdots , d_s\}$.\\
$D_+  r^\alpha=D_{\dot +}  \bar r^{\bar\alpha }= D_{\dot -}  s^{\tilde\alpha} =
D_-  \bar s^{\tilde{\bar\alpha} }=0$.

\vspace{.1cm}

Counting components, one finds that chiral and twisted chiral fields
contain as many components as a $(1,1)$ superfield, while semi-chiral
fields have twice as many, half of which turn out to be auxiliary.
This is not so surprising as chiral and twisted chiral superfields parametrize
$\ker (J_+ - J_-)$ and $\ker (J_++J_-)$ resp., where the $N=(2,2)$ algebra
closes off-shell. On the other hand, semi-chiral fields parametrize
$(\ker[J_+,J_-])^\perp$ where there is no off-shell closure and where
additional auxiliary fields should be introduced. Note that there is also a
complex
linear superfield $f$ satisfying the quadratic constraint $D_+D_-f=D_{\dot +}
D_{\dot -}\bar f=0$ \cite{book1} and a twisted complex linear
superfield $g$ defined by $D_+D_{\dot -}g= D_{\dot +}D_-\bar g=0$.
They can be represented in terms of semi-chiral fields, $f=r+\bar s$ and
$g=r'+s'$. Potentials which depend on semi-chiral fields through these
particular combinations were excluded in the analysis in \cite{icproc}, as
the resulting models exhibit gauge invariances. Aspects of semi-chiral
models with gauge invariances were studied in \cite{frans}.
Complex linear and twisted linear fields are alternatives to chiral and twisted
chiral fields resp. A detailed study of their properties is given in
\cite{mil}.

The action in $(2,2)$ superspace is
\begin{eqnarray}
{\cal S}=\int d^2x d^4\theta K(z,\bar z, w, \bar w,
 r,\bar r,  s,\bar s), \label{n2ac}
\end{eqnarray}
with $K$ a real potential. The potential is determined modulo a generalized
K\"ahler
transformation,
$K\simeq K+ f(z,w, r)+\bar f(\bar z, \bar w,\bar  r)+ g(z,\bar w,
\bar s)+\bar g(\bar z, w,  s)$.
Starting from eq. (\ref{n2ac}), one passes to $N=(1,1)$ superspace. Upon
elimination of the
auxiliary fields, one finds expressions for the metric,
torsion and complex structures in terms of the potential \cite{icproc}.
Various explicit examples are known.
K\"ahler manifolds are described using chiral fields only. The
$SU(2)\times U(1)$ Wess-Zumino-Witten (WZW) model can be described either in
terms of a
chiral and a twisted chiral multiplet \cite{RSS} or in terms of one semi-chiral
multiplet \cite{icproc,martinnew} depending on the choice one makes for
the complex structures. The WZW model on $SU(2)\times SU(2)$ is
described in terms of a semi-chiral and a chiral field \cite{icproc}.
Hyper-K\"ahler manifolds too can be described in terms of semi-chiral
coordinates \cite{icproc}. Indeed, choosing $J_+$ and $J_-$ such that
$\{J_+,J_-\}=0$, one gets $\ker [J_+, J_-]=\emptyset$.
\section{Ultra-violet properties}
Requiring conformal invariance severely restricts the allowed potentials
$K$ in eq. (\ref{n2ac}). The one loop $\beta$-function
of a general non-linear $\sigma$-model vanishes if
\cite{Rdil}, $R^+_{ab}+2\nabla^+_b\partial_a\Phi=0$, with $\Phi$ the
dilaton and $R^+_{ab}$ is the Ricci tensor comuted using the
$\Gamma_+$ connection. When only chiral and twisted chiral fields are present,
the
explicit expressions for metric and torsion, obtained from eq. (\ref{n2ac}),
are used in the analysis of the $\beta$-function \cite{kiritsis}.
Once semi-chiral fields enter the game, the expressions for
metric and torsion (see e.g. \cite{icproc} or \cite{BLR}) are so
complicated that this program becomes technically unfeasable. The way out
is to recompute the $\beta$-functions, but now directly in $(2,2)$
superspace. At present, there is no good understanding of the dilaton in
the presence of semi-chiral superfields, so we limit ourselves to the
study of a necessary condition for conformal invariance: ultra-violet
finiteness.

Computing the one-loop counterterms using the background field formalism
and the techniques described in \cite{book1} and \cite{book}, is quite
standard.
Full details are given in \cite{beta}.
We first consider a potential which only depends on  $d_s$ semi-chiral fields,
$K( r,\bar r, s ,\bar s)$.
The one-loop counterterm reads
\begin{eqnarray}
{\cal S}^{(1)}=\frac{ 1}{2\pi \varepsilon} \int d^2xd^4\theta\,
\ln\frac{\det {\cal N}_2}{\det (\sqrt{-3} {\cal N}_1)} , \label{1l}
\end{eqnarray}
where
\begin{eqnarray}
{\cal N}_1\equiv \left( \begin{array}{cc} K_{\widetilde\alpha \beta} &
K_{\widetilde\alpha \widetilde{ \bar\beta}}\\
K_{\bar\alpha \beta} & K_{\bar\alpha \widetilde{ \bar\beta}}\end{array}\right),
\qquad
{\cal N}_2\equiv \left( \begin{array}{cc} K_{\widetilde\alpha \bar\beta} &
K_{\widetilde\alpha  \widetilde{\bar\beta}}\\
K_{\alpha \bar\beta} & K_{\alpha
\widetilde{\bar\beta}}\end{array}\right).\label{ndef}
\end{eqnarray}
In \cite{icproc}, it was shown that non-degeneracy of the target manifold
metric is equivalent to $\det {\cal N}_1 \neq 0$ and  $\det {\cal N}_2 \neq 0$.
{}From eq. (\ref{1l}) we get the condition for UV finiteness at one loop:
\begin{eqnarray}
\det {\cal N}_2 = (-1)^{d_s}
|F( r)|^2 |G( s )|^2\det {\cal N}_1 , \label{d4con}
\end{eqnarray}
where $F$ and $G$ are arbitrary functions of $ r$ and $ s $ resp.
Note that there is no coordinate transformation compatible with the
constraints, which can remove $|F( r)|^2 |G( s )|^2$.

One checks this result by working through some
examples which are
known to be UV finite. The WZW model on $SU(2)\times U(1)$ is described by
a semi-chiral multiplet \cite{icproc,martinnew} with potential
\begin{eqnarray}
K=- (r+ \bar{ s})(\bar{ r} +  s)+\frac 1 2 (s+\bar s)^2
-2  \int^{s+\bar{ s }} dx \, \ln ( 1+ \exp \frac{x}{2}),
\label{su2u1}
\end{eqnarray}
and we find $F=1$ and $G=\exp(- s /2)$. Another class of interesting examples
are the
4-dimensional hyper-K\"ahler manifolds where $J_+$ and $J_-$ are chosen
to be anti-commuting. The potential then satisfies \cite{icproc}
$ |K_{ r s }|^2+|K_{ r\bar  s }|^2=2K_{ s \bar  s }K_{ r\bar r}$ so
that eq. (\ref{d4con}) is satisfied with $F=G=1$.

Finally, one can also repeat the calculation for a general potential, eq.
(\ref{n2ac}), which depends on all three types of superfields. One finds
the counterterm
\begin{eqnarray}
{\cal S}^{(1)}=\frac{ 1}{2\pi \varepsilon} \int d^2xd^4\theta\,
\ln\frac{\det (-K_{m\bar n})\det {\cal N}'_2}{\det K_{a\bar b}\det (\sqrt{-3}
{\cal N}'_1)} ,
\end{eqnarray}
where ${\cal N}'_1$ and ${\cal N}'_2$ are similar to ${\cal N}_1$ and ${\cal
N}_2$ in eq.
(\ref{ndef}) except that in ${\cal N}'_1$ one has to write the $d_s\times d_s$
matrices $(K_{AB}) -(K_{Aa})(K_{\bar a b})^{-1}(K_{\bar b B})$ instead of
the matrices $K_{AB}$, while in ${\cal N}'_2$ one has
$(K_{AB}) -(K_{Am})(K_{\bar m n})^{-1}(K_{\bar n B})$ instead of $K_{AB}$.
In these expressions,
the capital letters denote the indices appearing in eq. (\ref{ndef}),
while the small indices denote derivatives w.r.t. chiral or twisted chiral
fields, following the notation introduced in the definition of these
fields. Again, this result can be verified using a non-trivial example,
e.g. using the potential which describes the $SU(2)\times SU(2)$ WZW model in
terms of a semi-chiral and a chiral multiplet \cite{icproc}.
\section{Duality transformations}
The study of $T$-duality in $N=(2,2)$ superspace has the obvious advantage
that the extended supersymmetry remains manifest in the dual model.
Several aspects of dualities in $(2,2)$ superspace have been previously
studied. We refer the reader to \cite{dual} and the references therein. We
present
here a different approach starting from semi-chiral fields and imposing
further conditions on the prepotentials or assuming the existence of certain
isometries.

We begin with the first order lagrangian
\begin{eqnarray}
K(V,\bar V, W,\bar W,\cdots)-r V-\bar r\bar
V-s W-\bar s\bar W,\label{ord1}
\end{eqnarray}
where $V$ and $W$ are unconstrained complex prepotentials, $r$, $s$ are
semi-chiral fields and the dots stand for any kind of other superfields
which do not participate in the duality transformation. When no isometries are
present,
one gets four
dual formulations of the model, depending on whether one integrates over
$r$ and $s$, over $V$ and $W$, over $r$ and $W$ or over $s$ and $V$.

Imposing the constraint $V=\bar W$ in eq. (\ref{ord1}) yields $K(V,\bar
V)-f V-\bar f \bar V$, where we identified the complex linear field $f=r+\bar
s$. This is
the first order lagrangian for the well known duality which interchanges
a chiral for a complex linear superfield.
Alternatively, assuming the existence of an isometry such that
the potential in eq. (\ref{ord1}) depends only on the prepotentials
through the combination $V+\bar W$, we can rewrite eq. (\ref{ord1}) as
$K(V+\bar W,\bar V+W,\cdots)-\frac 1 2 (r+\bar s)( V+\bar W) -\frac 1 2
(\bar r+s )(\bar V+W) -\frac 1 2 (r-\bar s)( V-\bar W) -\frac 1 2
(\bar r-s )(\bar V-W)$. Integrating
over $V-\bar W$, requires that $r=\bar s$, which implies that $r$ is a
chiral field $z$. Calling $V'\equiv V+\bar W$, we get
the first order lagrangian  $K(V',\bar V',\cdots)-zV'-\bar z \bar V'$, which
makes it possible to pass from a complex linear field to a chiral field.
The twisted version of these dualities are obtained by interchanging the
roles of $W$ and $\bar W$. This time a twisted
chiral and a twisted complex linear superfield will be dual dual to each other.

Continuing to restrict our class of models, we assume an
isometry in the previously considered potential, $K(V+\bar
V)-f V-\bar f \bar V$, we perform the integration over $V-\bar V$ and
obtain that $f$ is real. Consistency of the constraints gives $f=\bar
f=w+\bar w$. Introducing the real prepotential $V'\equiv V+\bar V$, we get
the first order lagrangian  $K(V',\cdots)-(w+\bar w) V'$, which
makes it possible to pass from a chiral to a twisted chiral field. Both
descriptions have an Abelian isometry. On the other hand, the lagrangian
$K(V,\bar V,\cdots)-zV-\bar z \bar V$ allows for a further reality
constraint on the prepotential: $V=\bar V$ and we get $K(V,\cdots)-(z+\bar
z)V$.
The resulting model allows one to pass from a twisted chiral to a chiral
field.

Finally, the potential in eq. (\ref{ord1}) can allow for other constraints
or have more subtle isometries. Requiring that the prepotentials in eq.
(\ref{ord1}) satisfy $V+\bar V=W+\bar W$, we can parametrize the lagrangian by
$K(V'+W',\bar V'+\bar W',\bar V'+W',V'+\bar W',\cdots)-r(V'+W')-\bar r (\bar
V'+\bar W')
- s(\bar V'+ W')- \bar s(V' + \bar W')$. Integrating over the chiral fields
forces $V'$ to be chiral, $V'=z$, and $W'$ to be twisted chiral, $W'=w$.
Indeed the equations of motion for $r$ and $s$, $D_+(V'+W')=0$ and
$D_{\dot -}( \bar V'+W')=0$, imply that
$D_+D_{\dot -}(V'-\bar V')=D_{\dot +}D_{-}(V'-\bar V')=0$ and
$D_+D_-(W'-\bar W')=D_{\dot +}D_{\dot -}(W'-\bar W')=0$.
The resulting potential is given by $K(z+w,\bar z +\bar w, \bar z + w,\cdots )$
and exhibits an Abelian isometry, $z\rightarrow z+\varepsilon$,
$w\rightarrow w-\varepsilon$ with $\varepsilon$ a real constant.
Integrating first over the prepotentials, yields the dual model with a
potential of the form $K_{dual}(r-\bar r, r+\bar s,\bar r + s,\cdots)$,
which again has an Abelian isometry.

The inverse transformation can be obtained by requiring that the potential
in eq. (\ref{ord1}) has an isometry $K(V+\bar V, V+\bar W,\bar V + W,\cdots)$.
Integrating over the semi-chiral fields yields the model in terms of a
semi-chiral multiplet with potential, $K(r+\bar r , r+\bar s,\bar r
+s,\cdots)$. In order to obtain the dual model, we rewrite
eq. (\ref{ord1}) as $K(x, \bar u,u,\cdots)-x(r+\bar r - s - \bar
s)/2-iy(r-\bar r+s-\bar s)/2-u s-\bar u\bar s$,
where $x\equiv V+\bar V$,
$y\equiv -i(V-\bar V)$, $u\equiv \bar V +W$ and $\bar u \equiv V+\bar W$.
Integrating over $y$ requires that $r+s$ is real. Compatibility of this
with the constraints on $r$ and $s$ gives that we can express $r$ and $s$
in terms of a chiral and a twisted chiral field: $r=z+w$ and $s=\bar z-w$.
Subsequent integration over $x$, $u$ and $\bar u$ gives the dual potential
$K_{dual}(w+\bar w, \bar z-w, z-\bar w,\cdots)$.

Though a detailed discussion of the geometric properties of these duality
transformations will be given elsewhere, we do give here some explicit
examples. Particularly interesting is the $SU(2)\times U(1)$ WZW model.
Using the description in terms of a chiral and a twisted chiral field with
potential $K= \frac 1 2 (z + \bar z)^2-\int^{w+\bar w-z-\bar z}dx\ln
(1+e^x)$, we get by dualizing the twisted chiral field $K_{dual} = z'\bar
z' - \frac 1 2 (z+\bar z)^2-(z+\bar z) y -\int^y dx\ln(1+e^x)$, while
dualizing the chiral field yields $K'_{dual}=-w'\bar w' +\frac 1 2 (w+\bar
w)^2-(w+\bar w) y'+\int^{y'}dx\ln (1+e^x)$, where $y=\ln( e^{-z-\bar z}-1)$
and $y'=\ln(e^{w+\bar w}-1)$. In both cases, we obtain a free field and
the disk $SU(2)/U(1)$. The two descriptions of $SU(2)/U(1)$ are each
others dual.

Turning to the semi-chiral description of this model, eq. (\ref{su2u1}),
we notice that it has an Abelian isometry of the type previously
discussed. This allows us to dualize it to a chiral and a twisted field with
potential
\begin{eqnarray}
K_{dual}=z\bar z +2(\ln(1-e^{\frac{w+\bar
w}{2}}))^2-2\int^{w+\bar w-2\ln(1-e^{\frac{w+\bar w}{2}})} dx\ln (1+e^{\frac
x 2 }).
\end{eqnarray}
Again we notice a factorization of the dual model
into $(U(1))^2\times SU(2)/U(1)$.
\section{Conclusions}
The off-shell description of $N=(2,2)$ non-linear $\sigma$-models seems now
to be under control, though a precise geometric characterization of the
semi-chiral sector is still lacking. This would have particularly interesting
applications. E.g. as all hyper-K\"ahler manifolds allow
for a semi-chiral description, we obtain a potential, not the K\"ahler
potential, which allows for the computation of not only the metric, but all
three, mutually anti-commuting complex structures as well.

Our formulation reveals a rich
variety of duality transformations with the additional bonus that $(2,2)$
supersymmetry remains manifest. We summarize how the various duality
transformations act on superfields.
\begin{enumerate}
\item Dualities without isometries
\begin{itemize}
\item 4 dual formulations of a semi-chiral multiplet
\item complex linear $\leftrightarrow$ chiral
\item complex twisted linear $\leftrightarrow$ twisted chiral
\end{itemize}
\item Dualities with isometries
\begin{itemize}
\item chiral $\leftrightarrow$ twisted chiral
\item semi-chiral $\leftrightarrow$ 1 chiral + 1 twisted chiral
\end{itemize}
\end{enumerate}

The previous analysis was purely
classical. In order to get the quantum mechanical properties of these
duality transformations, a manifest $(2,2)$ supersymmetric description of
the dilaton coupling is needed. This necessitates a good understanding of
$N=(2,2)$ supergravity and its coupling to chiral, twisted chiral and
semi-chiral superfields. In \cite{DD}, $N=(2,2)$ supergravity was
investigated for backgrounds consisting of chiral and twisted chiral
superfields. It was found that the geometry of the $(2,2)$ super
worldsheet implies the existence of four types (anti-)chiral and twisted
(anti-)chiral) of worldsheet curvatures, which couple to fields
satisfying the same constraints. A generalization of this in the presence
of semi-chiral superfields is presently under study. Though the
ultra-violet properties of $(2,2)$ non-linear $\sigma$-models are well
understood, a good control over the dilaton would also enable one to
investigate the superconformal invariance of these models in detail.

\vspace{5mm}

\noindent {\bf Acknowledgments}:
We thank Silvia Penati, Martin Ro\v{c}ek and Daniela Zanon for useful
discussions.
M.T.G. is suported in part by the NSF Grant No. PHY-96-04587.
M.M., A.S. and J.T. are supported in part by the European
Commission TMR programme ERBFMRX-CT96-0045 in which all authors are associated
to K.U.Leuven.

\end{document}